\documentclass[prl,amsmath,twocolumn,amssymb,aps,superscriptaddress,floatfix,preprintnumbers,10pt,notitlepage]{revtex4}
\usepackage{times}
\usepackage{graphicx}
\usepackage{amsmath, braket, amsfonts}
\usepackage{amssymb}
\usepackage{natbib}
\usepackage{sidecap}
\usepackage{bm, color, ulem}
\usepackage{xcolor}
\usepackage{lipsum}
\usepackage{extarrows}
\usepackage{float}

\DeclareUnicodeCharacter{03BD}{{$\nu$}}

\begin{document}

\title{Strong-Magnetic-Field Magnon Transport in Monolayer Graphene}
\author{Haoxin Zhou}
\affiliation{Department of Physics, University of California at Santa Barbara, Santa Barbara CA 93106, USA}
\author{Chunli Huang}
\affiliation{Department of Physics, University of Texas at Austin, Austin TX 78712}
\author{Nemin Wei}
\affiliation{Department of Physics, University of Texas at Austin, Austin TX 78712}
\author{Takashi Taniguchi}
\affiliation{International Center for Materials Nanoarchitectonics,
National Institute for Materials Science,  1-1 Namiki, Tsukuba 305-0044, Japan}
\author{Kenji Watanabe}
\affiliation{Research Center for Functional Materials,
National Institute for Materials Science, 1-1 Namiki, Tsukuba 305-0044, Japan}
\author{Michael P. Zaletel}
\affiliation{Department of Physics, University of California, Berkeley, CA 94720, USA}
\affiliation{Materials Sciences Division, Lawrence Berkeley National Laboratory, Berkeley, California 94720, USA}
\author{Zlatko Papi\'c}
\affiliation{School of Physics and Astronomy, University of Leeds, Leeds LS2 9JT, United Kingdom}
\author{Allan H. MacDonald}
\affiliation{Department of Physics, University of Texas at Austin, Austin TX 78712}
\author{Andrea F. Young}
\email{andrea@physics.ucsb.edu}
\affiliation{Department of Physics, University of California at Santa Barbara, Santa Barbara CA 93106, USA}\date{\today}

\begin{abstract}
At high magnetic fields, monolayer graphene hosts competing phases distinguished by their breaking of the approximate SU(4) isospin symmetry. Recent experiments have observed an even denominator fractional quantum Hall state thought to be associated with a transition in the underlying isospin order from a spin-singlet charge density wave at low magnetic fields to an antiferromagnet at high magnetic fields, implying that a similar transition must occur at charge neutrality. However, this transition does not generate contrast in typical electrical transport or thermodynamic measurements and no direct evidence for it has been reported, despite theoretical interest arising from its potentially unconventional nature. Here, we measure the transmission of ferromagnetic magnons through the two dimensional bulk of clean monolayer graphene. Using spin polarized fractional quantum Hall states as a benchmark, we find that magnon transmission is controlled by the detailed properties of the low-momentum spin waves in the intervening Hall fluid, which is highly density dependent. Remarkably, as the system is driven into the antiferromagnetic regime, robust magnon transmission is restored across a wide range of filling factors consistent with Pauli blocking of fractional quantum hall spin-wave excitations and their replacement by conventional ferromagnetic magnons confined to the minority graphene sublattice.  Finally, using devices in which spin waves are launched directly into the insulating charge-neutral bulk, we directly detect the hidden phase transition between bulk insulating charge density wave and a canted antiferromagnetic phases at charge neutrality, completing the experimental map of broken-symmetry phases in monolayer graphene.
\end{abstract}
\maketitle 

\begin{figure*}
\centering
\includegraphics[width=\textwidth]{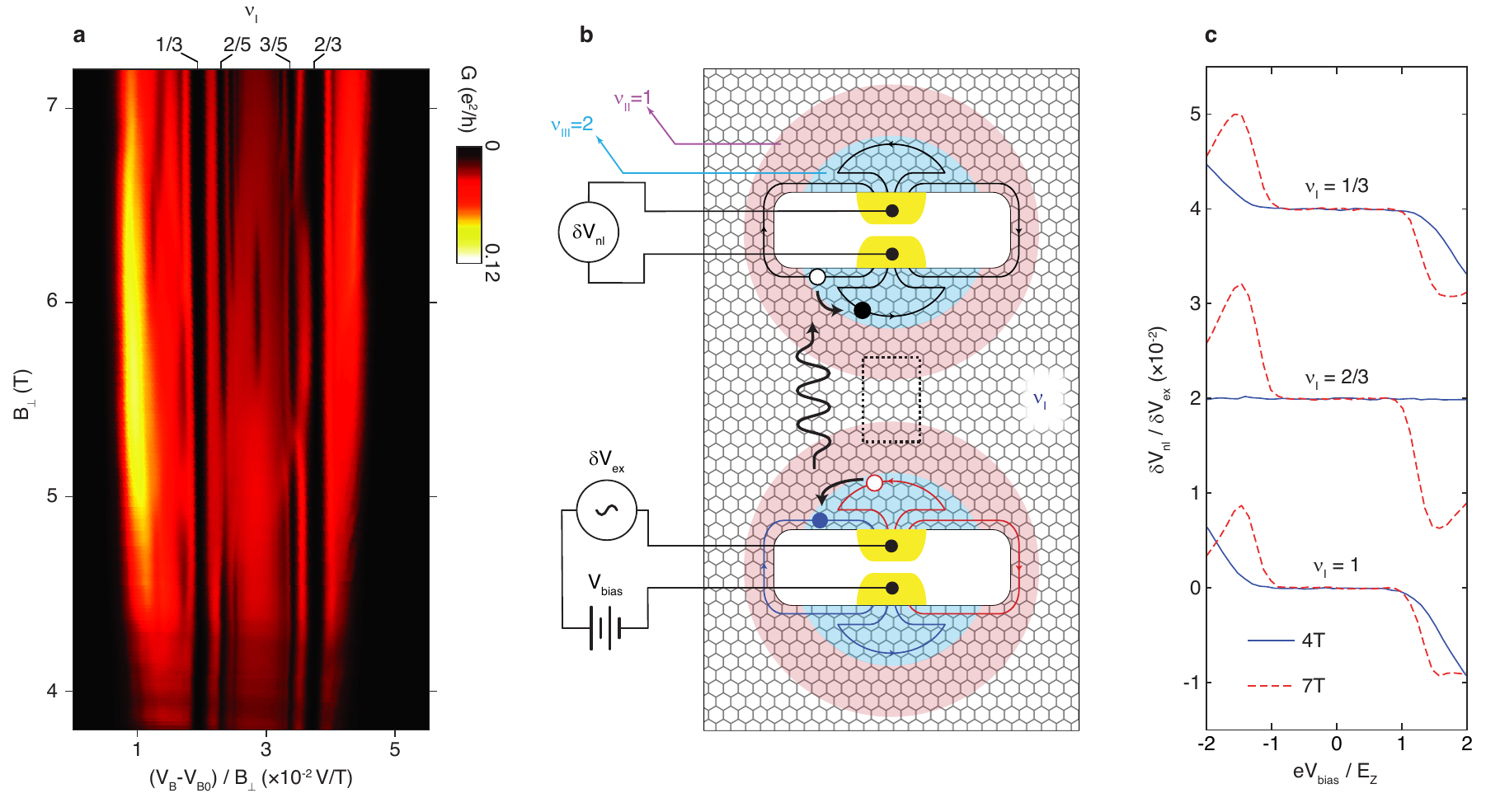} 
\caption{\textbf{Electrically detected magnon transmission in the fractional quantum Hall regime.}
\textbf{a}, Bulk conductance as function of perpendicular magnetic field ($B_\perp$) and gate voltage ($V_B$) showing fractional quantum Hall isospin transitions in the range corresponding Landau level filling factors $0<\nu_{\rm I}<1$.  The x-axis is offset and scaled by the applied magnetic field so that it is approximately proportional to $\nu_{\rm I}$.
\textbf{b}, Device schematic.  Electrostatic gates, fabricated from graphite, tune the filling factor $\nu_{\rm i}$ in three regions.  For $\nu_{\rm II}=1$ and $\nu_{\rm III}=2$, magnons can be generated by a bias voltage $ V_{\rm bias}$ 
and, if transmitted through region I, detected via a nonlocal electrical response $\delta V_{\rm nl}$ proportional to an applied AC excitation $\delta V_{\rm ex}$. 
\textbf{c}, nonlocal response as function of bias voltage $V_{\rm bias}$ for $\nu_{\rm I}=$1/3, 2/3 and 1 at $B_\perp=$4T and 7T. Curves are acquired at a nominal temperature of 70 mK, and each pair of curves is offset for clarity.
}\label{fig1}
\end{figure*}

Strongly interacting quantum magnets host a variety of spin and charge ordered states. 
In three dimensional materials, numerous probes are available that are directly sensitive to spin or charge order, allowing experiment to disambiguate competing states.
In two dimensional van der Waals heterostructures common bulk probes usually have insufficient sensitivity.  
In their place, experiment typically relies on electrical transport characterization, and on the ability to use electric and magnetic fields to tune microscopic parameters of the Hamiltonian.
Comparison with theoretical models can then be used to infer which phases are experimentally realized.  
Quantum Hall ferromagnetism in graphene provides a paradigmatic experimental example.  Here the intrinsic flatness of the Landau level bands makes Coulomb interactions dominant, while the spin- and valley degeneracy endow the Landau levels with a  multi-component nature that allows for a large number of competing orders.  
At charge neutrality, for example, predicted phases include a spin polarized ferromagnet (FM)\cite{fertig_luttinger_2006,abanin_spin-filtered_2006}, a canted antiferromagnet (CAF)\cite{herbut_theory_2007,jung_theory_2009,kharitonov_phase_2012}, a lattice scale charge density 
wave (CDW), and a partially sublattice polarized (PSP) bond-density wave\cite{nomura_field-induced_2009,kharitonov_phase_2012,kharitonov_canted_2012}.  
The variety of competing phases is even more abundant at nonzero Landau level filling\cite{abanin_fractional_2013,sodemann_broken_2014}, featuring a subtle interplay of ferromagnetic physics and the correlations underlying the fractional quantum Hall effect. 
Crucially, the relative favorability of different isospin-symmetry breaking  phases can be tuned experimentally by varying the charge carrier density, Zeeman energy and a substrate induced sublattice splitting. 

Experiments to date have focused on detection of the FM and CAF spin-ordered states, which are distinguished by edge mode properties that differentiate their two-terminal conductances\cite{young_tunable_2014, veyrat_helical_2020}.  
In most cases, however, phases cannot be distinguished by electrical transport.   
For example, transport cannot detect the transition at charge neutrality between spin-ordered FM or CAF states, and charge ordered CDW/PSP states, which has attracted theoretical attention as an analogue of the Ne\'el to valence bond solid transition studied in models of quantum magnetism\cite{lee_deconfined_2014,lee_wess-zumino-witten_2015}. 
This transition is expected to occur\cite{kharitonov_canted_2012} in samples with a finite sublattice splitting, which in monolayer graphene can be induced by a hexagonal boron nitride substrate\cite{hunt_massive_2013,amet_insulating_2013}.  
In this scenario, a CDW state obtains at low magnetic fields where the sublattice splitting is dominant. As the magnetic field is raised, the strength of the Coulomb interactions grows, including the strength of the short range interactions that distinguish the valleys and favor the CAF state. 
Once these are sufficiently large, the CDW gives way to the CAF phase via a partially sublattice polarized phase featuring a Kekul\'e distortion of the charge density wave order.  Experimentally, indirect evidence for this transition has been reported in samples that show a sublattice gap at B=0\cite{zibrov_even-denominator_2018}.  
The main electronic signatures of such a transition are illustrated in Figure \ref{fig1}a, which shows transport data measured in a Corbino geometry over a filling factor range $0<\nu<1$. Data are taken using Device A, 
which features a band gap of 3.7 meV (see Fig. \ref{figs_dv1}).
A series of phase transitions at odd denominator fillings are visible as local conductance maxima that mark points at which the bulk energy gap closes.  In addition, an even denominator state is observed in the neighborhood of $B_\perp^*\approx6$T. 
Taken together, these data were interpreted\cite{zibrov_even-denominator_2018} as signatures of a transition between underlying charge density wave to antiferromagnetic order, in which fractional occupation is transferred between the carbon sublattices.  However, no sign of the transition is observed in conventional transport at charge neutrality itself, where all the candidate states are electrical insulators, nor do typical electrical measurements give direct insight into the underlying isospin polarizations of the fractional states.

Here, we use electrically actuated magnon transmission measurements \cite{wei_electrical_2018,stepanov_long-distance_2018,zhou_solids_2019} to directly probe isospin polarization independent of its influence on transport. Our device geometry is shown in Fig.~\ref{fig1}b, and consists of an injector and detector separated by a region of arbitrary filling factor $\nu_{\rm I}$\cite{zhou_solids_2019}. 
The injector and detector each consist of a junction between regions of filling $\nu_{\rm II}$=1 and $\nu_{\rm III}$=2, and the device topology is such that no chiral edge states connect the injector and detector irrespective of the value of $\nu_{\rm I}$. 
Magnons are generated at the III-II interface in the injector by controlling the chemical potential 
difference $eV_{\rm bias}$ between co-propagating edge states of opposite spin. 
When the bias exceeds the threshold set by the Zeeman energy, $e V_{\rm bias}/E_{\rm Z}>1$, electrons can scatter between edge states conserving spin and energy by emitting a neutral magnon into the ferromagnetic $\nu_{\rm II}=1$ bulk. Here $E_{\rm{Z}} = g\mu_{\rm{B}}B$ is the Zeeman energy, $g=2$, $\mu_{\rm B}$ is the Bohr magneton, and $B$ the total applied magnetic field.
When region I also supports long-lived neutral modes that couple strongly to region II magnons, 
spin and energy can be transmitted across region I to the detector.  Within the detector, absorbtion of a magnon generates a voltage that can be detected via the nonlocal response $\delta V_{\rm nl}/\delta V_{\rm ex}$, where $\delta V_{\rm nl}$  is a finite-frequency nonlocal 
voltage induced at the detector in response to a small excitation $\delta V_{\rm ex}$ applied to the injector.   

\begin{figure*}
\centering
\includegraphics[width=\textwidth]{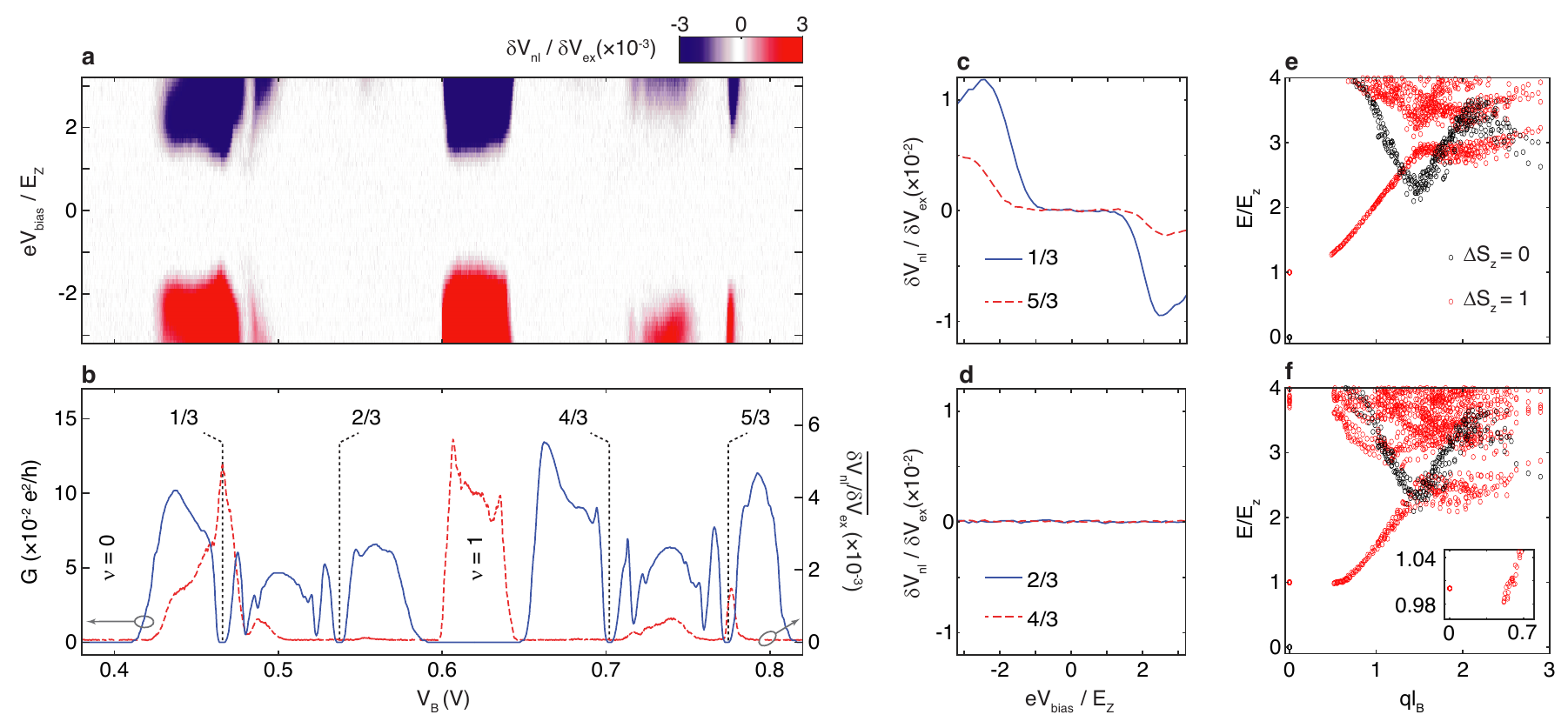}
\caption{
\textbf{Contrasting magnon transmission in spin-polarized fractional quantum Hall states.}
\textbf{a}, Nonlocal response as function of the bottom gate voltage $V_{\rm B}$ and $eV_{\rm bias}/E_{\rm Z}$ for $0<\nu_{\rm I}<2$ measured at $B=4$T and nominal temperature $T=50$mK.
\textbf{b}, Comparison of the root-mean-square average of the data in panel b (red, dashed) and bulk conductance (blue, solid) as a function of $V_{\rm B}$.
\textbf{c}, Nonlocal response at $\nu_{\rm I}=$ 1/3, 5/3 and 
\textbf{d}, $\nu_{\rm I}=$ 2/3, 4/3 as a function of $eV_{\rm bias}/E_{\rm Z}$, extracted from data in panel \textbf{a}.
\textbf{e}, Spectrum of neutral modes with spin $\Delta S_z=0$ and $1$ measured relative to the ground state for $\nu=1/3$ at $B_\perp=4$T calculated using exact diagonalization techniques (see Methods).
\textbf{f}, The same as panel e, but for the the spin-polarized 2/3 state. Inset: zoomed-in near $E/E_{\rm Z}=1$.
}\label{fig2}
\end{figure*}

Fig. \ref{fig1}c shows $\delta V_{\rm nl}/\delta V_{\rm ex}$ measured at $\nu_{\rm I}=1$, 1/3, and 2/3 at magnetic fields both above and below $B_\perp^*$. 
At $\nu_{\rm I}=1$, region I is density matched to the injector and detector, resulting in a large nonlocal response at or slightly above the Zeeman threshold for all $B$\cite{wei_electrical_2018,zhou_solids_2019} corresponding to ferromagnetic magnon transmission through the uniform $\nu_{\rm I} = 1$ bulk. 
Similarly, the nonlocal signal is strong both above and below $B_\perp^*$ at $\nu_{\rm I}=1/3$, indicating magnon propagation.  However, magnon transmission is undetectable at $\nu_{\rm I}=2/3$ for $B=4$T but is restored at 7T, indicating a change in the neutral mode spectrum associated with the transport phase transition. 
We focus first on the nonlocal signal for $B<B_\perp^*$, shown in Fig.\ref{fig2}a-b. 
Besides $\nu_{\rm I}=1$, the strongest nonlocal response is observed at $\nu_{\rm I}=5/3$ and over a range $.12<\nu_{\rm I}<.4$ that includes $\nu_{\rm I}=1/3$; in contrast, no nonlocal response is observed for even numerator fractional quantum Hall states states ($2/3$ and $4/3$). 

It is tempting to ascribe the observed even-odd effect to a difference in ground state spin polarization; after all, in the absence of a Zeeman energy the 2/3 and 4/3 states are expected to be spin-singlet, and as a consequence should not host propagating spin wave excitations\cite{halperin_theory_1983}. 
However, tilted field magnetotransport measurements\cite{polshyn_quantitative_2018} indicate that the applied $B_\perp=4$T is sufficient to spin polarize these states (see Fig.~\ref{figs_tf}). 
For all spin polarized states, Larmor's theorem dictates that the spin wave spectrum features at least one mode whose energy increases quadratically as $E(q)\approx E_{\rm Z}+\hbar^2 q^2/2m$ and whose  lifetime diverges as $q \to 0$, with the $q=0$ Larmor mode corresponding to a uniform rotation of all spins. 
Consequently, the suppression of magnon transmission for even numerator $n/3$ states cannot arise from an \textit{absence} of spin wave excitations; rather, it indicates that the specific  properties of the spin excitations at fractional filling are incompatible with transmission of ferromagnetic magnons from the $\nu=1$ quantum Hall ferromagnet. This constitutes a much stronger constraint on the ground state wave function and its excitations than simply sharing a non-zero spin polarization.

\begin{figure*}[t!]
\centering
\includegraphics[width=\textwidth]{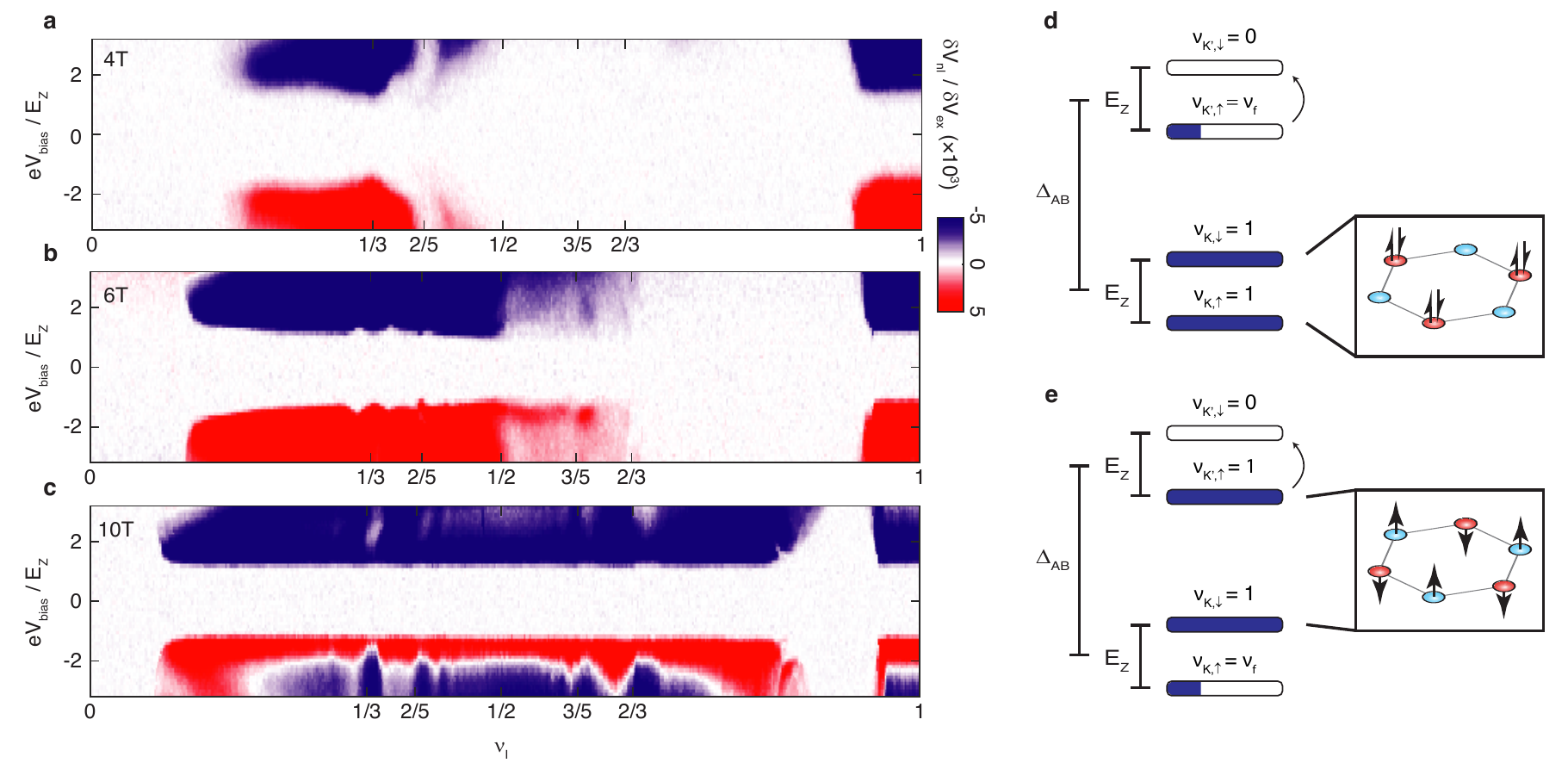}
\caption{\textbf{Magnon transmission at a fractional quantum Hall isospin transition.}
\textbf{a}, Nonlocal response measured for $0<\nu_{\rm I}<1$ as a function of $eV_{\rm bias}/E_{\rm Z}$ at $B_{\perp}=$ 4T. $\nu_{\rm I}$ is calibrated using the bulk conductivity (see Fig. \ref{figs_nl}).
\textbf{b}, The same as panel a at $B_{\perp}=6$T.
\textbf{c}, The same as panels a and b at $B_{\perp}=10$T.
\textbf{d}, Single particle energy level diagram for the zero Landau level in the low magnetic field regime.  The $K$ valley is fully occupied, corresponding to a sublattice polarized state (inset).  Additional fractional occupation is in the $K',\uparrow$ level.  
Spin-1 excitations with $E(q=0)=E_{\rm Z}$ consist of excitations between the partially filled and empty $K'$ states, indicated by the arrow. 
\textbf{e}, The same diagram but for the high-field phase. Now the underlying order is antiferromagnetic (inset).  Spin-1 excitations from the fractionally occupied level are Pauli blocked, with low energy excitations possible only between the filled $K',\uparrow$ and empty $K',\downarrow$ levels--identical to those of the quantum Hall ferromagnets in the injector and detector. 
}\label{fig3}
\end{figure*}

For magnon transmission to be observed region I must host long-lived spin-wave excitations that have  
have significant overlap with the magnons of the $\nu=1$ quantum Hall ferromagnet over a range of wave vectors, permitting high transmission at the I/II interface, propagation to the second II/I interface and eventual absorption into the detector.  
While there is no general theoretical framework to calculate the full spin-wave dispersion for a partially filled Landau level, exact diagonalization of the interacting Hamiltonian is possible at $\nu=1/3$ and $2/3$ and  neutral mode spectra calculated for monolayer graphene are shown in Figs.\ref{fig2}e-f (see Methods). 
At $\nu=1/3$, the spin-waves are consistent with a simple evolution of the Larmor mode into a single, monotonically increasing spin wave branch that eventually merges with a spin-flip continuum for $E/E_{\rm Z}\approx 3$.
In contrast, calculations at $\nu=2/3$ produce a qualitatively different spin wave dispersion, in which the mode energy at the lowest finite value of $q$ accessible to the numerical calculations is \textit{lower} than the energy at q=0. 
This behavior is consistent with a finite momentum `spin-roton minimum' in the dispersion, as previously reported based on both numerical calculations~\cite{mandal_low-energy_2001,majumder_neutral_2014} and inelastic light scattering measurements~\cite{wurstbauer_observation_2011}. 
Physically, it arises from the interplay between the strong Zeeman effect, which polarizes the ground state, and Coulomb interactions which favors an unpolarized state\cite{halperin_theory_1983}. Finite momentum spin waves correspond to modulations of the ground-state spin density that locally lower the spin polarization and thus the Coulomb energy, leading to a spin-roton minimum at finite-$q$ (see supplementary information).
At $\nu=2/3$, the spin-flip continuum also appears at lower energy, closer to $E/E_{\rm Z}\approx 2$. 

The calculated spectra suggest several possible explanations for the suppression of magnon transmission at $\nu=2/3$ (and $4/3$).  First, the flatter spin-wave dispersion may decrease the transmission of magnons across the I/II interface due to kinematic constraints, in a magnetic analog of Kapitza resistance. Second, magnons with $E>E_{\rm Z}$ may decay inelastically into the spin roton minimum; there, they have energy $E<E_{\rm Z}$ and consequently cannot enter the detector. Finally, the presence of a lower lying spin-flip continuum, combined with disorder\cite{kallin_many-body_1985}, may 
provide a damping channel at $\nu=2/3$ that is not present at $\nu=1/3$.
Indeed transport measurements suggest that the bottom of the spin-flip continuum lies at $\Delta\sim1.5E_{\rm Z}$
for $\nu=2/3$ and $4/3$ in similar devices \cite{polshyn_quantitative_2018}.
While we are not able distinguish between these mechanisms in our current experiment---and indeed, they may all be relevant even at a single filling factor---we note that all are expected to be highly sensitive to the details of the ground state wave-function and its spin-flip excitations.  At the most qualitative level, this is consistent with experiment where an intricate dependence on $\nu_{\rm I}$ is observed both for $0<\nu_{\rm I}<1$ when $B<B_\perp^*$ and for $1<\nu_{\rm I}<2$ at all magnetic fields (see Fig. \ref{fig:evenodd}).

Remarkably, this intricate dependence vanishes at $B>B_\perp^*$, replaced by a density independent restoration of magnon transmission across a a continuous range of $\nu_{\rm I}$.  
As shown in Fig.~\ref{fig3}a-c, for $B>B_\perp^*$ the nonlocal signal onsets sharply at the Zeeman threshold---indicating high magnon transmission--throughout the $0<\nu<1$ range with the exception of narrow regions near $\nu=0$ and 1 where signatures of electron solids have been observed\cite{zhou_solids_2019}.  
The $B_\perp$-dependent change in transmission cannot be attributed to a change in the ground state spin polarization, as follows from absence of any dependence of  $B_\perp^*$ (or the magnetic fields corresponding to gap closings at odd denominator fillings) on the in-plane magnetic field (see Fig.~\ref{figs_tft} and Ref.~\onlinecite{zibrov_even-denominator_2018}). 
Above $B_\perp^*$ the magnon transmission becomes completely insensitive to the $\nu-$dependent details of the spin-wave dispersion, in apparent conflict with our previous observation that magnon transmission is strongly modulated by the detailed spin-wave spectra of fractional quantum Hall states. 

We understand the decoupling of magnon transport from the fractional quantum Hall effect as a consequence of the isospin phase transition thought to underlie the incompressible state at $\nu=\pm1/2$. 
Figs. \ref{fig3}d-e show the expected isospin polarization above and below $B_\perp^*$. 
In both cases, electrons occupy three of the four isospin flavors, leaving the fourth empty. 
Below $B_\perp^*$, the underlying order is that of the CDW state so the low-energy carbon sublattice corresponding to the majority valley $K$ is occupied with both spin projections, $\nu_{K\uparrow}=\nu_{K\downarrow}=1$. The remaining electrons occupy the available minority sublattice states with the Zeeman favored spin projection, $\nu_{K'\uparrow}=\nu$. Since all available states in the majority valley are completely occupied, the excitation spectrum projected onto the minority valley is identical to a two-component quantum Hall system \cite{moon_spontaneous_1995}, leading to the situation described in Fig. \ref{fig2}. 

Above $B_\perp^*$, in contrast, the underlying order is that of the AF state so $\nu_{K'\uparrow}=\nu_{K\downarrow}=1$ and the remaining electrons occupy the favored sublattice with the Zeeman favored spin-projection $\nu_{K\uparrow}=\nu$. 
Both spin and sublattice isospins are active.
In this case, spin waves in the minority valley involve a transition from a completely occupied Landau level ($\nu_{K'\uparrow}=1$) to an empty Landau level ($\nu_{K'\downarrow}=0$). As a result, they are expected to closely resemble excitations within the $\nu=1$ injector and detector regions.  
Indeed, within a double-mode approximation that accounts for collective spin-lowering transitions on both sublattices (see supplementary information), 
we find that the spin-wave stiffness increases only slightly compared to that of the magnons at $\nu=1$.  This is compatible with high transmission of magnons between these ground states, and the observed high nonlocal signal.
The onset of nonlocal response across the entire Landau level is thus a direct signature of the transition to antiferromagnetic order.  
This interpretation is further supported by the contrasting behavior of the nonlocal response for $1<\nu_{\rm I}<2$  (Fig. \ref{figs_nl}), where antiferromagnetism is not expected and where no such ubiquitous response is observed.

Despite direct evidence for an AF-CDW transition at $\nu\neq 0$, magnon transmission is not observed near the Zeeman threshold at charge neutrality for any magnetic field. 
The absence of a signal at low fields is expected, since the CDW phase is nonmagnetic.  However, in the high field phase, a noncollinear canted antiferromagnet is thought to be the ground state because the Zeeman energy favors spin canting when the two sublattices have 
equal occupation\cite{kharitonov_phase_2012,abanin_fractional_2013}. 
Although prior work\cite{wei_electrical_2018,stepanov_long-distance_2018} has reported nonlocal response at $\nu=0$, it 
has been observed only at bias voltages far above the Zeeman threshold, where for example Joule heating by the injector is non-negligible\cite{zhou_solids_2019}. Theoretical calculation within the linear spin-wave approximation \cite{wei_scattering_2020} suggests the absence of a low-energy nonlocal signal is because of the mismatch between the magnon dispersions of $\nu=0$ CAF and the $\nu=1$ FM. Additionally, the presence of two magnon modes in the CAF opens up magnon-magnon and other decay channels.

\begin{figure}[t!]
\centering
\includegraphics[width=\columnwidth]{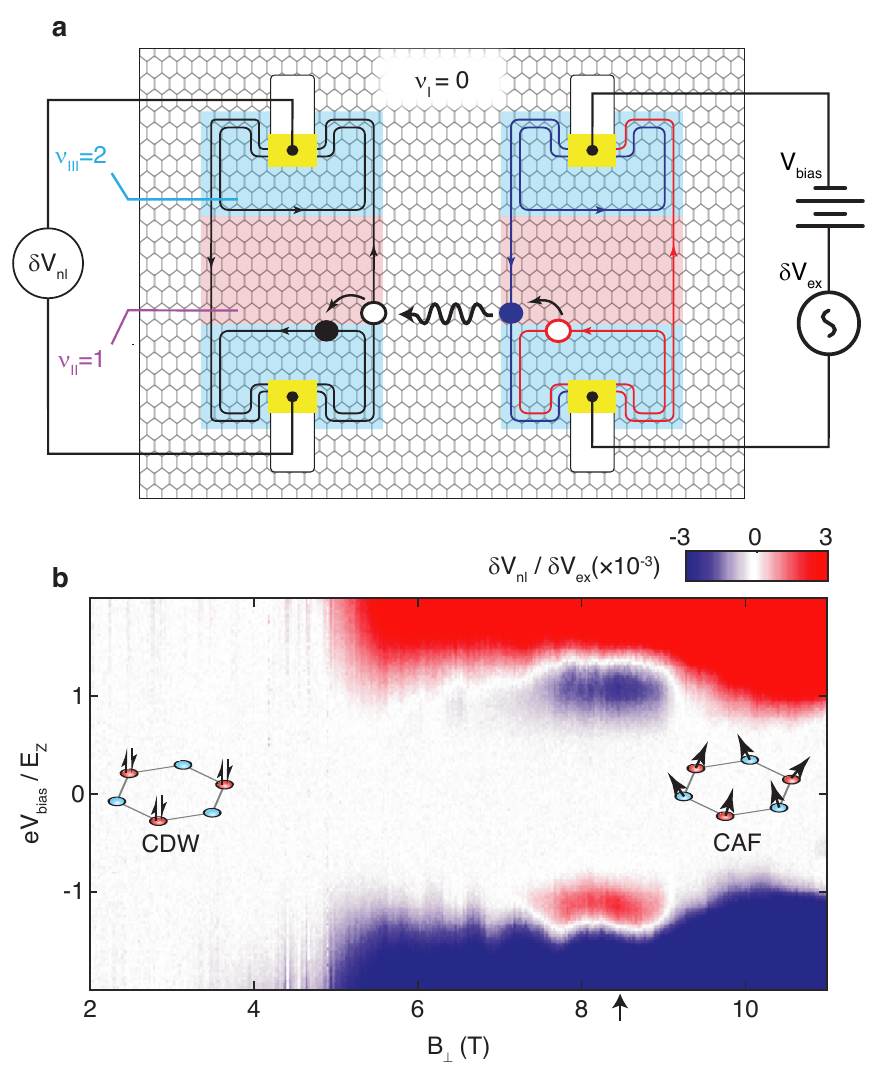}
\caption{
\textbf{Charge density wave to canted antiferromagnet transition at ${\bm \nu=0}$.}
\textbf{a}, Schematic of the second experimental geometry.  Regions rendered in white, red and blue are gate-defined regions on the monolayer graphene, yellow regions are metal contacts.
\textbf{b}, nonlocal response vs $eV_{\rm bias} / E_{\rm Z}$ and $B_{\perp}$ at $\nu_{\rm I}=$ 0. The arrow indicates the value of $B_\perp$ where the incompressible $\nu = 1/2$ state appears.
}\label{fig4}
\end{figure}

To disambiguate these issues and more directly probe the magnon transmission at $\nu=0$, we introduce a different sample 
geometry shown in Fig. \ref{fig4}a. 
In this scheme, spin wave generation and detection follow the same mechanism, with the contact configuration allowing for a potential imbalance between co-propagating $\nu=1$ and $\nu=2$ edge states.  In contrast to the device shown in Fig. \ref{fig1}a, in the second geometry, magnons may be generated directly at a boundary with the $\nu=0$ state, and nonlocal response need not be mediated by $\nu=1$ magnons\cite{takei_spin_2016}. 
Fig. \ref{fig4}b shows the measured nonlocal response in Device B, which shows an even denominator fractional quantum Hall state at $B_\perp^*=8T$ (see Fig. \ref{figs_dv2}). A clear onset of nonlocal response at the Zeeman threshold is observed around $B_{\perp}=$ 5T, indicating a phase transition from a non-magnetic phase to a magnetic phase. 
With an in-plane field applied, the phase transition is shifted to a smaller $B_{\perp}$ (Fig. \ref{figs_tfn0}). 
This is in excellent agreement with the expectations for the CAF phase, which is favored by the Zeeman energy as compared to the nonmagnetic CDW phase due to its canted spin structure and finite magnetic moment\cite{kharitonov_phase_2012,abanin_fractional_2013}. 

Our detection of the CDW-AF transition confirms the theoretically predicted picture of isospin symmetry broken phases in monolayer graphene at charge neutrality. However, we note that our data also raises a number of new questions.  While we have focused on the most tractable filling factors of $\nu=0$ and integer multiples of 1/3, the magnon data of Figs. \ref{fig1}b, \ref{fig2}b-d, and additional data available in the supplementary material show a highly featured evolution of the magnon transmission with fractional filling, the precise mechanisms of which remain to be understood.  A more full theoretical interpretation of these features may provide insight into their spin- and valley polarizations, as well as the nature of neutral modes in strongly interacting systems more generally.  
An additional intriguing feature of the data is the qualitative change in the nonlocal signal in Sample B in the neighborhood of the $\nu=1/2$ state shown in Fig. \ref{fig4}b. 
Absent a quantitative theory of magnon transmission, this signature is difficult to interpret, but may shed light on the 
nature of the correlated state, which is currently not well understood. 
Finally, a mean-field analysis of the phase diagram of the $\nu=0$ state predicts the existence of a Kekul\'e distorted PSP phase between the CDW and CAF phases. Future experiments with sensitivity to the sublattice polarization, and theoretical treatments of the magnon transmission in this regime, may shed light on this enigmatic phase. 

\paragraph{Methods}
Devices were fabricated using a dry transfer procedure. Sample A was fabricated following refs. \cite{zeng_high-quality_2019, polshyn_quantitative_2018}. Details of the heterostructure are shown in Fig. \ref{figs_dv1}(a). Sample B was fabricated by first assembling and patterning a heterostructure containing a monolayer graphene and two graphite gates, shown in Fig. \ref{figs_dv2}(a), and then subsequently transferring third graphite gate. In both of the devices, hexagonal boron nitride flakes are used to isolate conducting layers, which are not shown in Fig. \ref{figs_dv1}(a) and \ref{figs_dv2}(a).

All data except that in Fig. \ref{figs_nl}m and n were acquired in a dilution refrigerator equipped with a 14 T superconducting magnet. The measurements were performed at base temperature unless indicated, corresponding to a measured temperature of T$\lesssim$30 mK on the probe. Data in Fig. \ref{figs_nl}m and n were performed in a dilution refrigerator equipped with a 18 T superconducting magnet at a base temperature $T\lesssim$ 50 mK indicated by the thermometer on the probe. The local conductance and nonlocal voltage measurements on Sample A require tuning of the carrier density in region II and III, following Ref. \onlinecite{zhou_solids_2019}. The differential conductance was measured using a lock-in amplifier with a 100 $\mu$V excitation at 17.777 Hz. The nonlocal voltage was measured using a lock-in amplifier with an ac excitation at 1234.5 Hz with various amplitudes on the order of 100 $\mu$V. The frequency is chosen to reduce the noise while maintaining negligible phase shift, and the amplitude is chosen to balance the signal to noise ratio and sampling precision. The nonlocal voltage measurements on Sample B were performed with an ac excitation of 99.2$\mu$V at 17.777 Hz.

Exact diagonalization calculations of the neutral mode spectra were performed using the torus geometry~\cite{chakraborty_fractional_1987}. This geometry, which does not suffer from the ambiguity of ``shift"~\cite{wen_shift_1992}, allows us to obtain the exact low-lying energy spectrum at filling factor $\nu$, resolved as a function of a two-dimensional momentum $\mathbf{q}$~\cite{haldane_many-particle_1985} and the total $z$-projection of spin $S_z$ of the particles. We diagonalized systems of $N=6-10$ electrons (holes) at filling factor $\nu=1/3$ ($\nu=2/3$), with up to two spin-flips away from the maximal polarization ($S_z \geq N/2-2$). In order to obtain finer resolution of the collective mode, we collated the data corresponding to different types of unit cells, varying the angle between the lattice vectors going from square to hexagonal unit cell, with aspect ratio of the torus fixed to unity. The effective interaction potential is taken from Ref.~\cite{yang_experimental_2020}, and includes the dielectric constant $\epsilon_{\mathrm{hBN}}=\sqrt{\epsilon^{||}\epsilon^\perp}$, with $\epsilon^\perp=3.0$ and  $\epsilon^{||}= 6.6$, as well as the  screening  by  the  graphite  gates, which are accounted for using standard electrostatic calculations, and by the filled Dirac sea at the RPA level~\cite{shizuya_electromagnetic_2007}.

\begin{acknowledgments}
AFY and HZ acknowledge discussions with I. Sodemann.  
AHM CH and NW acknowledge support from the ARO under Grant Number W911NF-16-1-0472 and 
from the Welch Foundation under grant F1473.  
Z.P. acknowledges support by the Leverhulme Trust Research Leadership Award RL-2019-015.
MPZ acknowledges support from the ARO through the MURI program (grant number W911NF-17-1-0323).
Experimental work by H.Z. and A.F.Y. was supported by the National Science Foundation under DMR-1654186. 
A portion of this work was performed at the National High Magnetic Field Laboratory, which is supported by the National Science Foundation Cooperative Agreement No. DMR-1644779 and the state of Florida. 
K.W. and T.T. acknowledge support from the Elemental Strategy Initiative conducted by the MEXT, Japan, Grant Number JPMXP0112101001,  JSPS KAKENHI Grant Number JP20H00354 and the CREST(JPMJCR15F3), JST.
\end{acknowledgments}
\normalem

\ULforem


\renewcommand\thefigure{S\arabic{figure}}
\setcounter{figure}{0}

\begin{figure*}
\centering
\includegraphics[width=\textwidth]{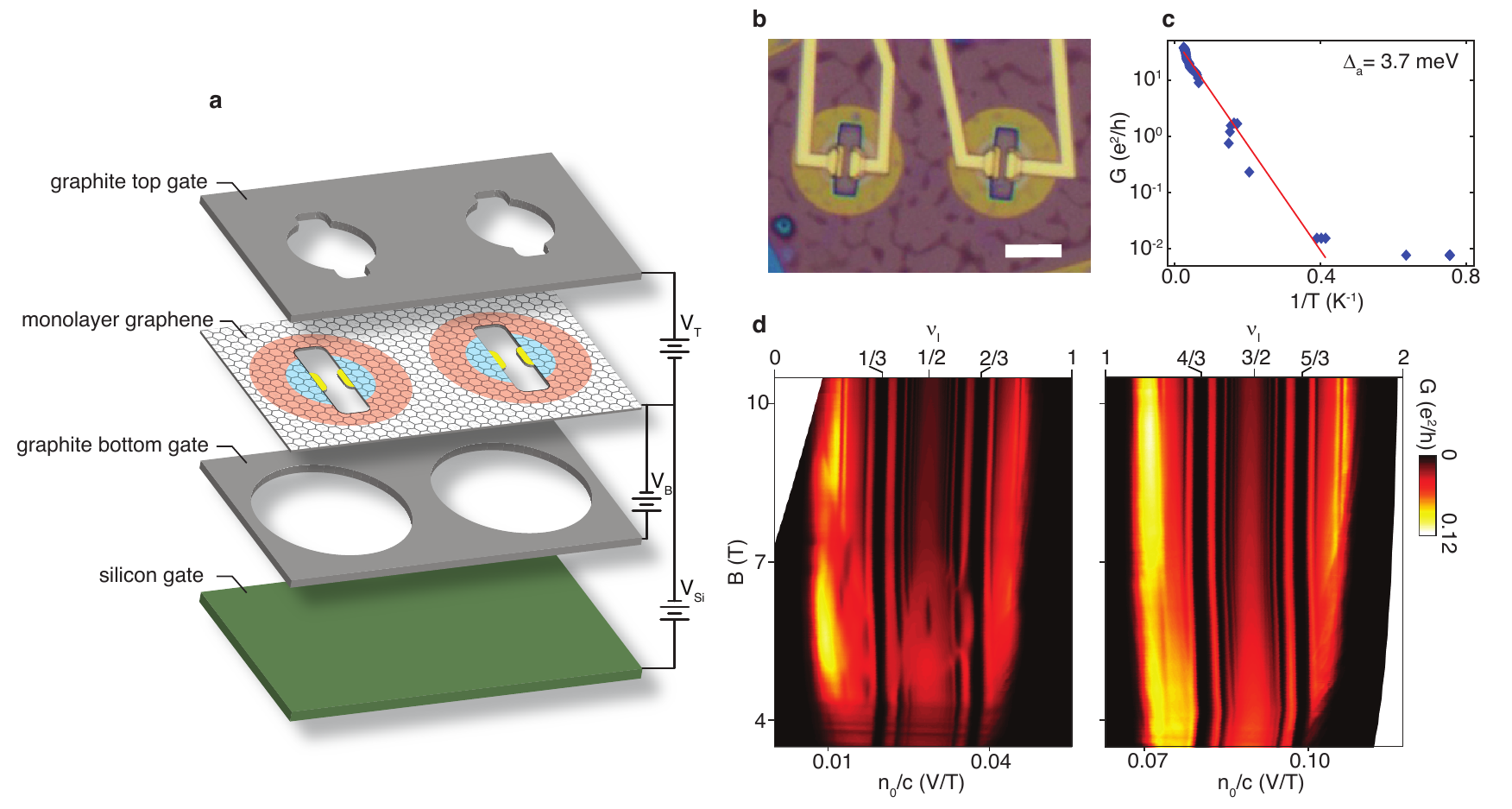} 
\caption{\textbf{Geometry and transport characterization of Sample A.}
\textbf{a}, Device geometry.
\textbf{b}, Optical micrograph of the sample. Scale bar represents 5 um.
\textbf{c},Arrhenius plot of the conductance at the charge neutrality point, linear fitting of the data shows the sample is insulating with an energy gap of 3.7 meV.
\textbf{d},Conductance versus carrier density and magnetic field between $\nu =$0 to $\nu =$2. Phase transitions are only observed between 0 $<\nu<$ 1 but not 1 $<\nu<$ 2, consistent with refs. \cite{zibrov_even-denominator_2018} and \cite{polshyn_quantitative_2018} in the main text.
}\label{figs_dv1}
\end{figure*}

\begin{figure*}
\centering
\includegraphics[width=\textwidth]{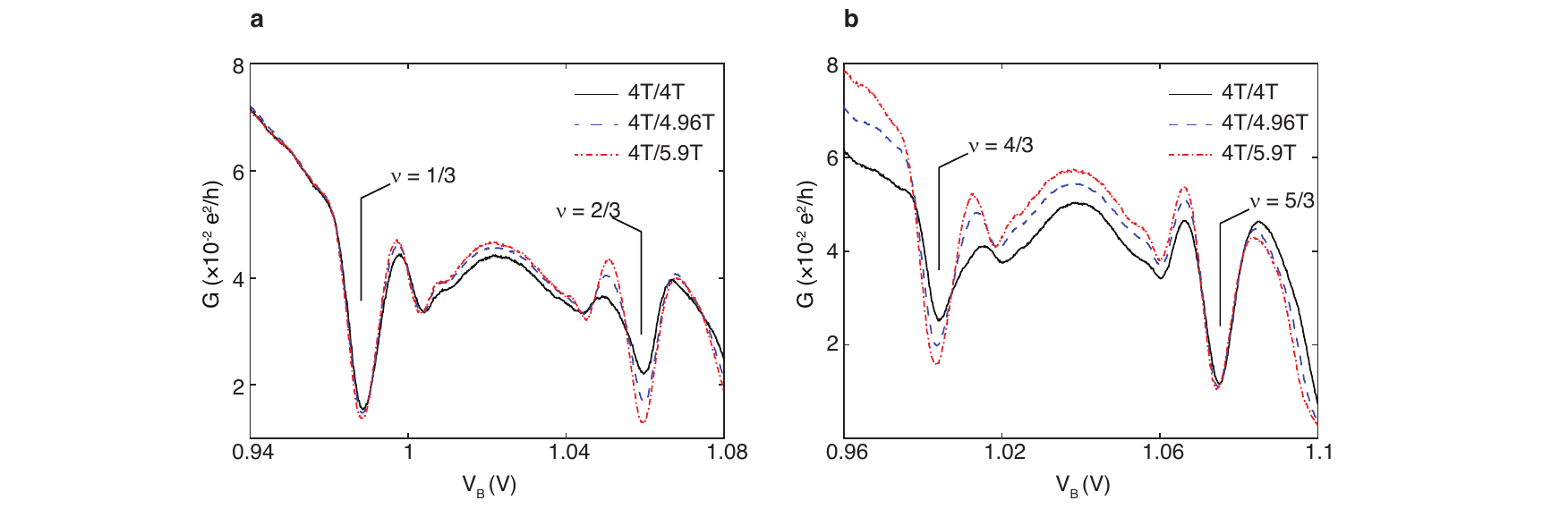} 
\caption{\textbf{In-plane field dependence of thermally activated conductance of fractional quantum Hall states in the zeroth Landau level.} Measurement was performed at T = 1.6K, where the $\nu=$ 1/3, 2/3, 4/3 and 5/3 states are well-activated. $B_{\perp}$ is fixed at 4T. The data shows significant decrease of the conductance at $\nu=$2/3 and 1/3 as the in-plane magnetic field increases. The conductance at $\nu=$ 1/3 and 5/3, however, are not sensitive to in-plane magnetic field. The result is consistent with ref. \cite{polshyn_quantitative_2018} in the main text, indicating the $\nu=$ 2/3 and 4/3 states are single-component fractional quantum Hall states.}\label{figs_tf}
\end{figure*}

\begin{figure*}
\centering
\includegraphics[width=120mm]{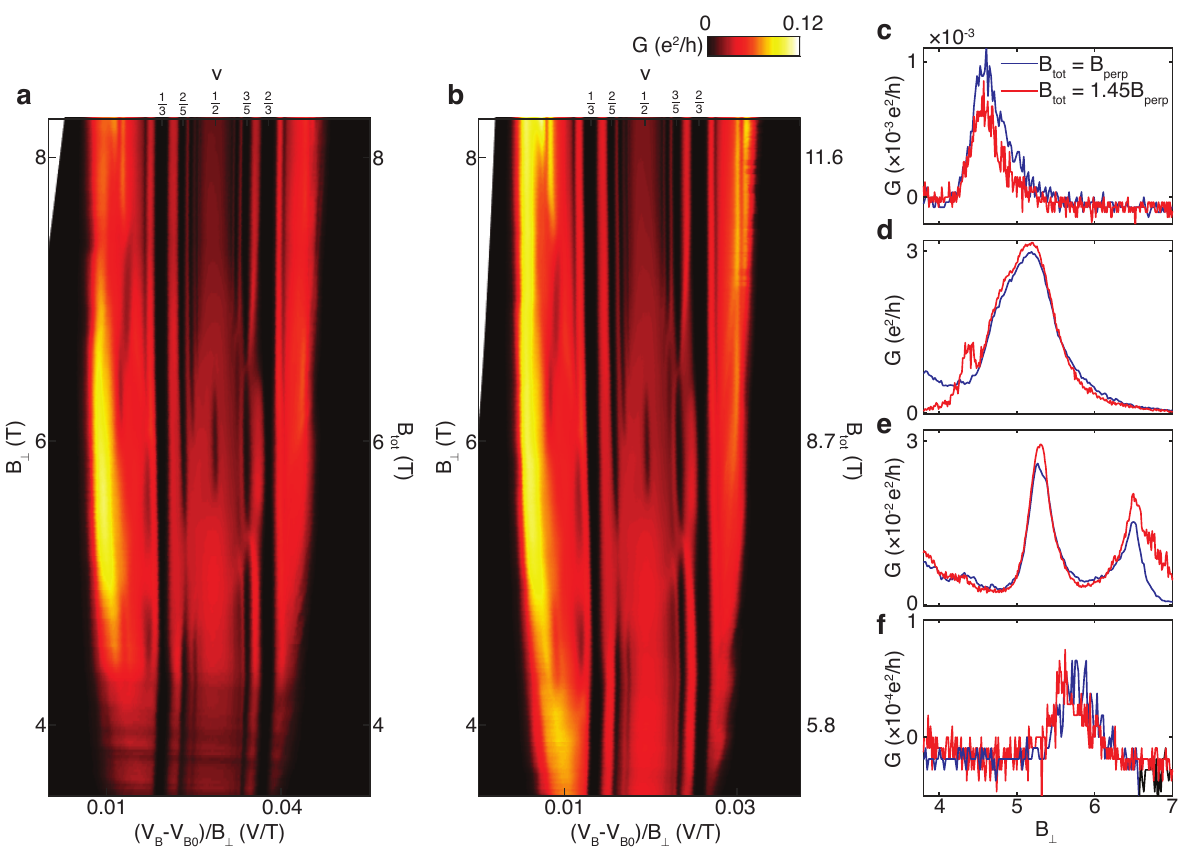} 
\caption{\textbf{Tilted field transport measurement on Sample A}
\textbf{(a)}, Local conductance $G$ as a function of carrier density and out-of-plane magnetic field.
\textbf{(b)}, Same as panel a with an in-plane magnetic field applied. The ratio of $B_{\rm tot}$ and $B_{\perp}$ is fixed at 1.45.
\textbf{c-f}, Vertical line-cuts of panel a and b at $\nu =$ 1/3, 2/5, 3/5 and 2/3. The local maxima of $G$ indicate the phase transitions. The line-cuts clearly show that these transitions do not depend on the in-plane magnetic field.
}\label{figs_tft}
\end{figure*}

\begin{figure*}
\centering
\includegraphics[width=\textwidth]{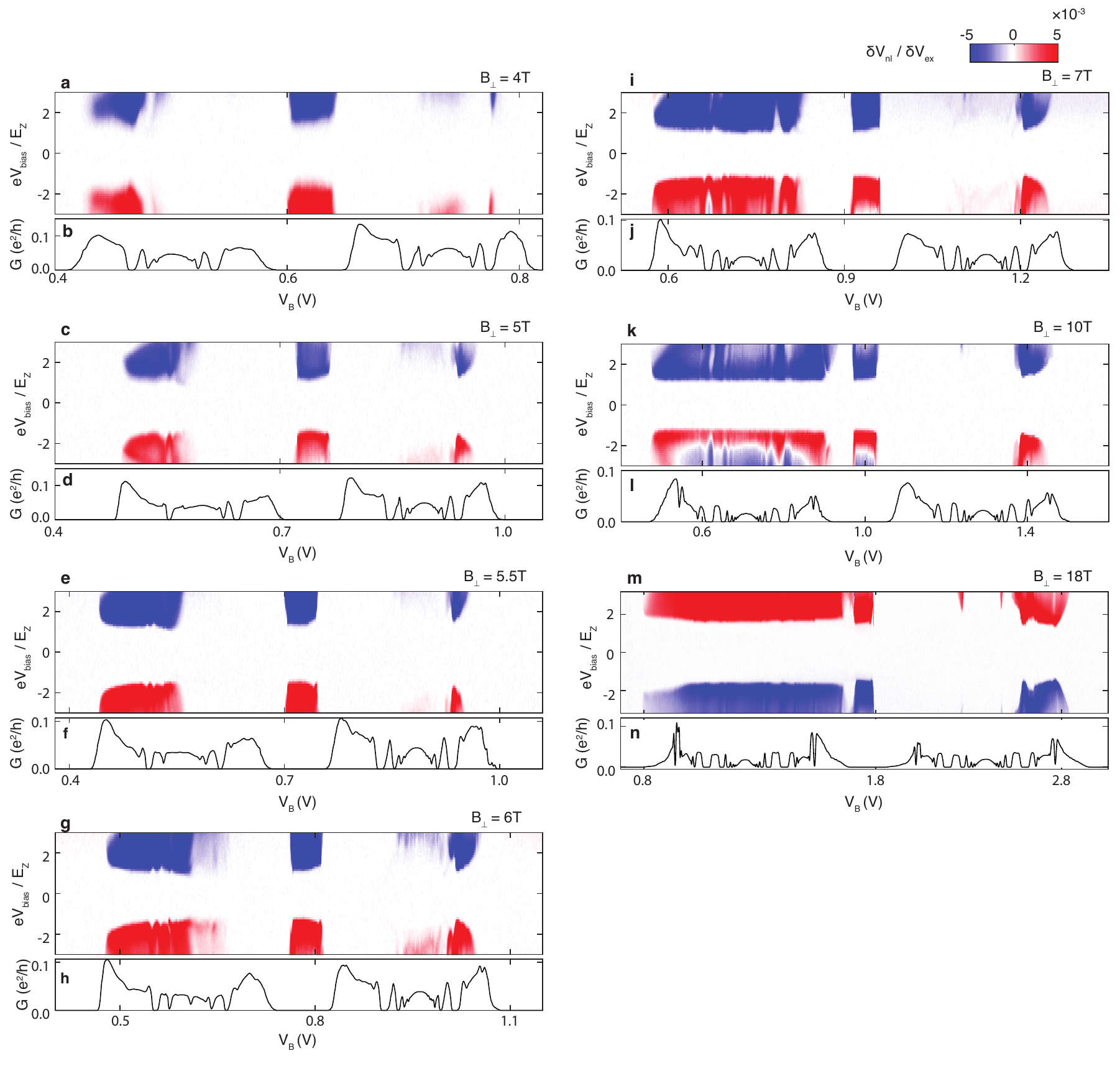} 
\caption{\textbf{Nonlocal response and local conductance between ${\bm \nu=0}$ and ${\bf \nu=2}$ at different values of ${\bm B_\perp}$.} The local conductance is plotted in the same range as the nonlocal voltage, acting as a calibration of the carrier density. 
}\label{figs_nl}
\end{figure*}

\begin{figure*}
\centering
\includegraphics[width=120mm]{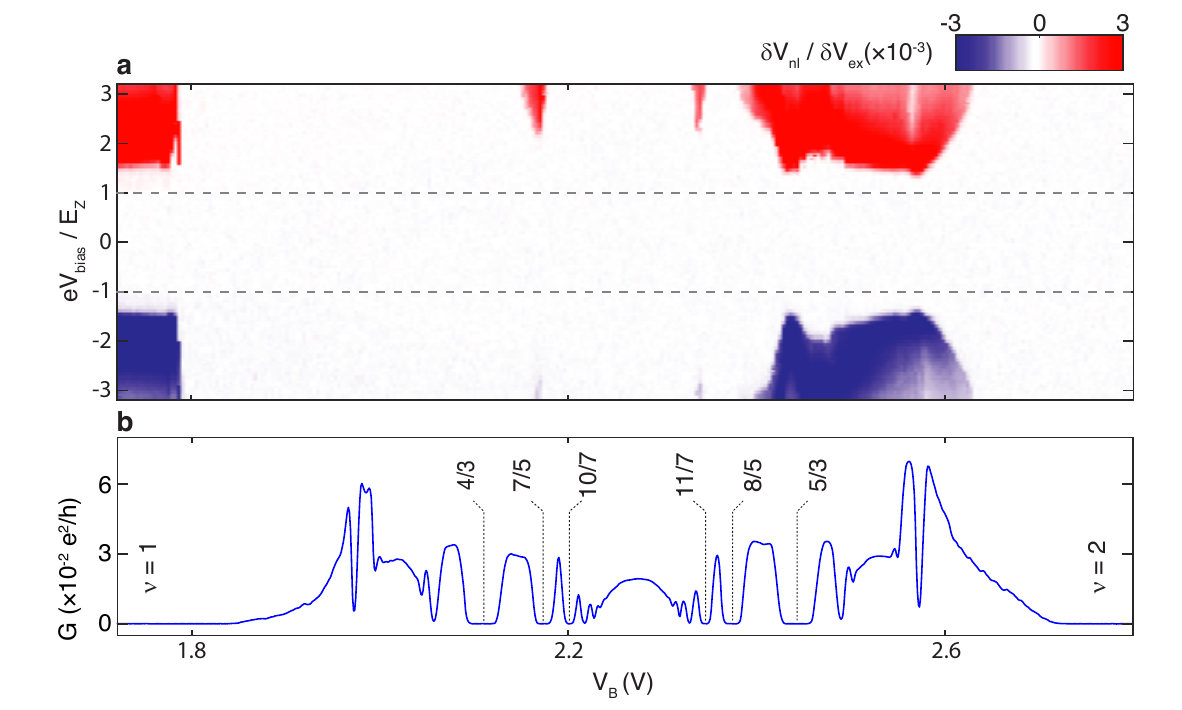} 
\caption{\textbf{Expanded view of nonlocal response between ${\bm 0<\nu_{\rm I}<1}$ at ${\bm B_{\perp}=18}$T of Sample A}
\textbf{(a)}, nonlocal response as function of $V_{\rm B}$ and $\delta V_{\rm bias}/\delta V_{\rm ex}$ for  $1<\nu_{\rm I}<2$  measured at $B=$18T and nominal T=70 mK. 
\textbf{(b)}, bulk conductance as a function of $V_{\rm B}$. 
}\label{fig:evenodd}
\end{figure*}

\begin{figure*}
\centering
\includegraphics[width=\textwidth]{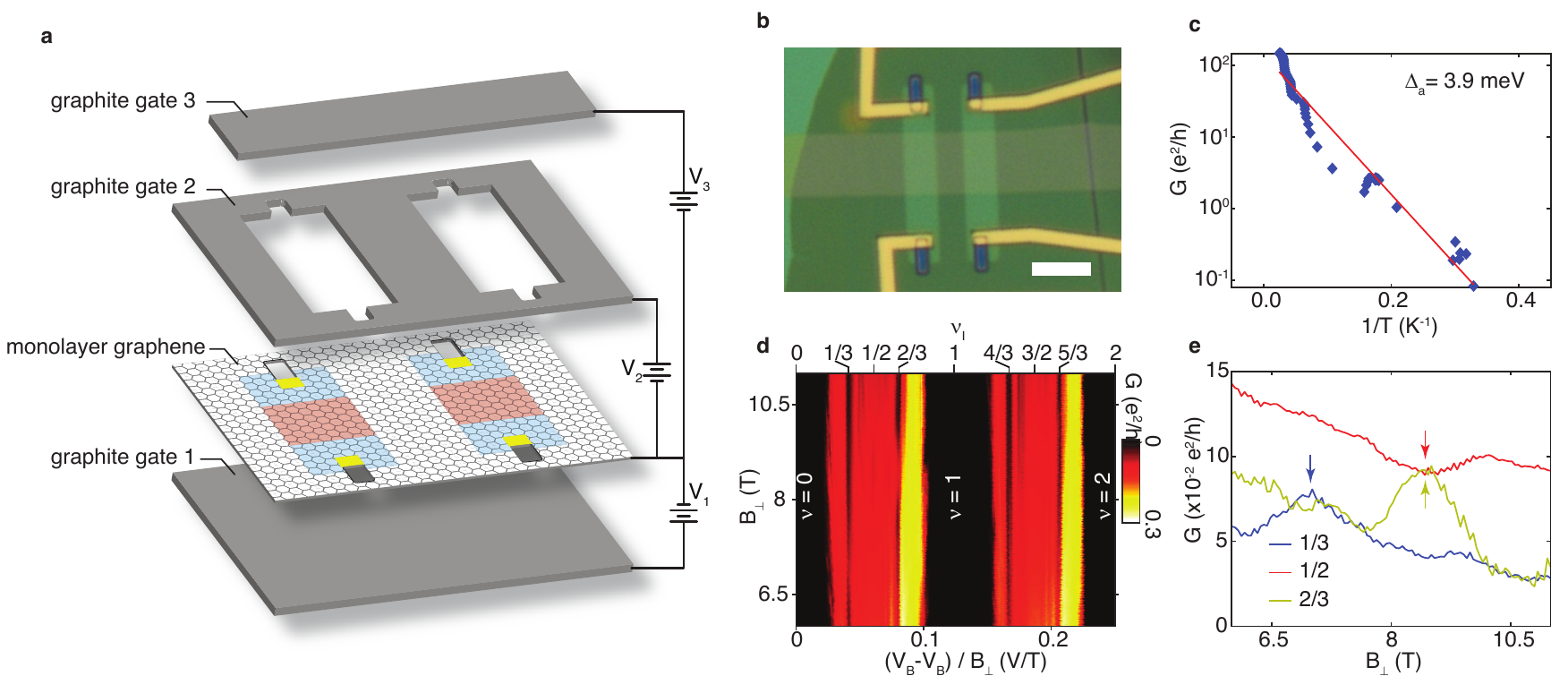} 
\caption{\textbf{Geometry and transport characterization of Sample B.}
\textbf{(a)}, Device geometry.
\textbf{(b)}, Optical micrograph of the sample. Scale bar represents 5 um.
\textbf{(c)}, Arrhenius plot of the conductance at the charge neutrality point, linear fitting of the data shows the sample is insulating with an energy gap of 3.9 meV.
\textbf{(d)}, Conductance versus carrier density and magnetic field between $\nu =$0 to $\nu =$2.
\textbf{(e)}, Line-cuts of panel c along vertical axis at $\nu=$ 1/3, 1/2 and 2/3. Phase transitions at $\nu=$ 1/3, 2/3 and the incompressible state at $\nu=1/2$ are marked by arrows.
}\label{figs_dv2}
\end{figure*}

\begin{figure*}
\centering
\includegraphics[width=120mm]{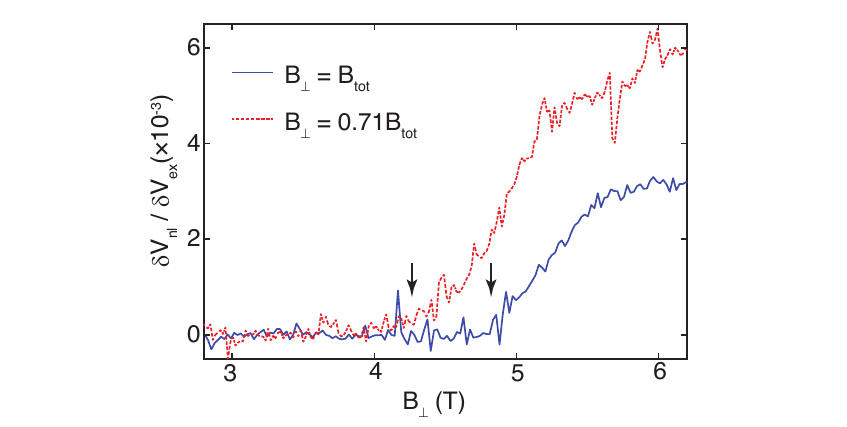} 
\caption{\textbf{In-plane field dependence of the phase transition at $\nu=$ 0.}
Solid line: Line-cut of Fig. \ref{fig4}\textbf{b} at $eV_{\rm bias} / E_{\rm Z} = 1.74$. Dashed line: Same measurement with a tilted magnetic field applied. $B_{\perp} = 0.71B_{\rm tot}$. The arrows label the value of $B_{\perp}$ where the nonlocal voltage starts to increase.
}\label{figs_tfn0}
\end{figure*}

\newpage
\clearpage

\section{Magnon transmission: Mean-field approach}

In this section, we discuss magnon transmission for a device geometry consists of an integer quantum Hall ferromagnet with filling faction $\nu_{\rm II}=1$ and a state at filling fraction $0<\nu_{\rm I}<1$.
Ref.~\cite{wei_scattering_2020} presents microscopic calculations of the magnon transmission probability for the case of $\nu_{\rm I}=0$ and $\nu_{\rm I}=-1$ using time-dependent Hartree-Fock theory. The result is that the average magnon transmission for $\nu_{\rm I}=0$ canted antiferromagnet (CAF) is lower than $\nu_{\rm I}=-1$ ferromagnet (FM). This is because the magnon dispersion of the CAF is stiffer than FM so the energy and transverse-momentum conservation limits the available phase-space for transmission.
This effect resembles that underlying Kapitza heat resistance\cite{khalatnikov_introduction_2018}. 
Based on the idea of energy mismatch, we proceed to give a rough estimate of magnon-transmission probability. We assume long-wavelength magnon dispersion in $\nu_{\rm I}$ region is parameterized by a density dependent spin-stiffness $\rho_\nu$: $\hbar\omega_{\nu}(q)=E_{\rm Z}+\rho_{\nu}q^2$. Because of the quadratic dispersion, a magnon's equation of motion satisfies the Schr\"odinger equation and we solve it via the elementary means:

\begin{align}\label{eq_ansatz}
    \psi(x)&=Ae^{iq_Lx}+Be^{-iq_{L}x},\ \ x<0\notag\\
    &=Ce^{iq_{R}x},\qquad\qquad x>0
\end{align}
where $\rho_{1}q_{L}^2=\rho_{\nu_{\rm I}}q_{R}^2$. Current conservation and wave function continuation imposes the following boundary conditions:
\begin{align}\label{eq_bc}
    \rho_{1}q_L(A-B)=\rho_{\nu_{\rm I}}q_R \; \; ,\;\;
    A+B=C.
\end{align}
By solving the above equations, we obtain the ratio of the transmitted and injected spin current as follows
\begin{equation}\label{eq_T}
    T=\frac{j_{trans}}{j_{tot}}=\frac{\rho_{\nu_{\rm I}}q_{R}|C|^2}{\rho_{1}q_{L}|A|^2}=\frac{4\sqrt{\rho_{1}\rho_{\nu_{\rm I}}}}{(\sqrt{\rho_{1}}+\sqrt{\rho_{\nu_{\rm I}}})^2}
\end{equation}

We shall now estimate the magnon dispersion for the weak-field CDW-like state and the strong field AF-like state.


\subsection{Magnon transmission at \texorpdfstring{$B<B_\perp^{*}$}{low field}}

For the weak-field state $\ket{\Psi_{L}}$, the majority valley is fully occupied with both spin-projections while the minority fractionally occupied the Zeeman favored spin-projection. As a result, the ground state has the following property:
\begin{equation}\label{eq_groundstate}
    s_{i}^{+}\ket{\Psi_{L}}=\tau_{i}^{+}\ket{\Psi_{L}}=0,\quad\forall, i=1,2...N_{e}
\end{equation}
where $N_e$ is the number of electrons where $s_i^+$ and $\tau_i^+$  are, respectively, the spin and valley creation operators.


The zero energy Landau level projected spin lowering operator is given by the following operator:
\begin{align}
S_{q,B}^{-}=\sum_{j}e^{i\vec{q}\cdot\vec{R}_{j}}\frac{1-\tau_{j}^{z}}{2}s_{j}^{-}.
\end{align}
This operator acts on the guiding center coordinate $\vec{R}$ of all the electrons and create a single magnon in minority valley ($K'$) with wave vector $\vec{q}$.

The single-mode approximation (SMA) assumes the spectral weight of spin-1 excitations at wavevector $\vec{q}$ is mainly concentrated at the energy $\hbar \omega_v(\vec{q})$ which is the energy required to create $S_{q,B}^{-}|\psi_L\rangle$. This approximation is exact at $q=0$ since it satisfies Larmor's theorem and it remains a good approximation in the long-wavelength limit.

Starting from the Heisenberg equation of motion, we arrive at the familiar  equation for the SMA:
\begin{equation}
    \omega_{\nu}(\vec{q})= \frac{\langle [S_{-q,B}^{+},[H,S_{q,B}^{-}]]\rangle}{\langle S_{-q,B}^{+}S_{q,B}^{-}\rangle}
    \label{eq_sma}
\end{equation}

where $\langle\rangle$ denotes the expectation value with respect to $\ket{\Psi_{L}}$. The Hamiltonian projected to the zero Landau level is given by the following:
\begin{align}
    &H=H_{0}+H_{z}+H_{\perp}-\frac{\Delta_{AB}}{2}\tau_{k=0}^z - \frac{\Delta_z}{2}\, S_{k=0}^z\\
    &H_{0}=\frac{1}{2A}\sum_{k}g_{0}(k)\rho_{-k}\rho_{k}\\
    &H_{z}=\frac{1}{2A}\sum_{k}g_{z}(k)\tau_{-k}^{z}\tau_{k}^{z}\\
    &H_{\perp}=\frac{1}{4A}\sum_{k}g_{\perp}(k)(\tau_{-k}^{+}\tau_{k}^{-}+\tau_{-k}^{-}\tau_{k}^{+})
\end{align}
Here we defined $\rho_{k}=\sum_{j}e^{i\vec{k}\cdot\vec{R}_{j}}$ and $\Gamma_{k}=\sum_{j}e^{i\vec{k}\cdot\vec{R}_{j}}\Gamma_{j}$ for arbitrary combination $\Gamma$ of Pauli matrices.
\begin{align}
g_{0}(k)=\frac{e^{2}}{\epsilon}\frac{2\pi}{k}e^{-\frac{k^{2}l_{B}^{2}}{2}},\qquad g_{z/\perp}(k)=g_{z/\perp}e^{-\frac{k^{2}l_{B}^{2}}{2}}
\end{align}
Using the commutation relation $[R_{i}^{\mu},R_{j}^{\nu}]=-il_{B}^2\epsilon^{\mu\nu}\delta_{ij},(\mu,\nu=\{x,y\})$, we can evaluate the following double commutators for the later convenience:
\begin{widetext}
\begin{equation}\label{eq_commutators0z}
    [S_{-q,\alpha}^{+},[H_{0}+H_{z},S_{q,\alpha}^{-}]]=\frac{4}{A}\sum_{k}(g_{0}(k)+g_{z}(k))\sin^{2}\left(\frac{k\wedge q}{2}\right)\bigg(\frac{1}{2}\{S_{-k-q,\alpha}^{+},S_{k+q,\alpha}^{-}\}-\rho_{-k}\rho_{k}\bigg),
\end{equation}
\begin{align}\label{eq_commutatorsxy}
[S_{-q,\beta}^{+},[H_{\perp},S_{q,\alpha}^{-}]] & =\frac{c_{\alpha}c_{\beta}}{2}\sum_{k}g_{\perp}(k)e^{\frac{i}{2}\vec{k}\wedge\vec{q}(c_{\alpha}-c_{\beta})}\times\bigg(\big\{(\tau^{+}s^{+})_{-k-q},(\tau^{-}s^{-})_{k+q}\big\}\nonumber \\
 & +\big\{(\tau^{-}s^{+})_{k-q},(\tau^{+}s^{-})_{k+q}\big\}-4\big\{(\tau^{-}\frac{1-s^{z}}{2})_{k},\tau_{-k}^{+}\big\}-4\big\{(\tau^{+}\frac{1+s^{z}}{2})_{-k},\tau_{k}^{-}\big\}\bigg)
\end{align}
\end{widetext}
where $\alpha,\beta=\{A,B\}$ and $c_{\alpha}=1(-1)$ for $\alpha=A(B)$. In this subsection, we are only interested in $\alpha=\beta=B$. Substituting the above two equations into the Eq.~\eqref{eq_sma}, applying Eq.~\eqref{eq_groundstate} to simplify the expression and using the spin susceptibilty,
\begin{equation}
    \langle S_{-q,B}^{+}S_{q,B}^{-}\rangle=4N_{B\uparrow},
\end{equation}
where $N_{s,\alpha}$ is the electron number on sublattice $\alpha$ with spin $s$, we derive the magnon dispersion and spin stiffness,
\begin{align}
\omega_{\nu}(\vec{q})&=E_{z}+\frac{2}{A}\sum_{k}\big[g_{o}(k)+g_{z}(k)\big]\sin^{2}\left(\frac{k\wedge q}{2}\right) \notag \\
&\;\; \times \left(1-\mathcal{S}_{\nu}(k)\right)
,\label{eq_omegaq}\\
\rho_{\nu}&=\frac{1}{4}\int\frac{d^2k}{(2\pi)^2}\big[g_{o}(k)+g_{z}(k)\big]k^2\left(1-\mathcal{S}_{\nu}(k)\right). \label{eq:rho_nu}
\end{align}
Here $\mathcal{S}_{\nu}(k)=N_{B\uparrow}^{-1}\langle \rho_{-k}\rho_{k}\rangle$ is the static structure factor. It has the following properties:
 1) $\mathcal{S}_{\nu}(k)$ is only non-vanishing in the fractionally
occupied Landau level, i.e.~$\mathcal{S}_{1}(k)=0$ and 2) $\mathcal{S}_{\nu}(k)$ satisfies the following equation under particle-hole transformation, $\nu\mathcal{S}_{\nu}(k)=(1-\nu)\mathcal{S}_{1-\nu}(k)$ for $k\neq 0$.
Using these properties and Eq.~\eqref{eq_omegaq}, we arrived at the following equation relating the mode frequencies at different $\nu$:
\begin{equation} 
\omega_{\nu}=\left( 2-\frac{1}{\nu} \right) \omega_{1}+\left( \frac{1}{\nu}-1 \right)\omega_{1-\nu}.\label{eq_dispersionph}
\end{equation}

For $\nu=1$, the spin stiffness can be easily calculated from Eq.~\ref{eq:rho_nu} which yields 
$\rho_{1}=\sqrt{2\pi}e^2/8\epsilon l_B+u_{z}/2$. For $\nu=\frac{1}{3}$, we use the static structure factor evaluated in Ref.~\onlinecite{girvin_magneto-roton_1986} to calculate the spin stiffness $\rho_{1/3}$. When the ground state of $\nu=2/3$ is assumed to be the particle-hole conjugation of the fully spin-polarized $\nu=1/3$ Laughlin state, we can use Eq.~\eqref{eq_dispersionph} to estimate the magnon excitation energy. This lead to the following result:
\begin{align} 
    &\rho_{\frac{1}{3}}\approx 0.035\frac{e^2}{\epsilon l_{B}}\xlongequal[]{B=4T}0.1\rho_{1},\\ &\rho_{\frac{2}{3}}=\frac{1}{2}(\rho_{1}+\rho_{\frac{1}{3}})\xlongequal[]{B=4T}0.55\rho_{1}. \label{eq:stiffness}
\end{align}

The SMA predicts the magnon excitation energy at finite $q$ is always larger than the Larmor mode at $q=0$. When we substitute their dispersions into Eq.~\eqref{eq_T} to estimate the transmission probability, we arrived at
\begin{equation}
    T_{\frac{1}{3}}=73\%,\qquad T_{\frac{2}{3}}=98\%.
\end{equation}
For $\nu=1/3$, the SMA is in good agreement with finite-size exact-diagonalization (ED) calculation \cite{rezayi_reversed-spin_1987} for small $q$. For $\nu=2/3$, the SMA fails to produce the spin-roton minimum revealed by ED (as shown in Fig. 2 of main text and described inthe associated discussion). 
Evidently, substantial spin-flip spectral weight is placed in higher energy states that are not captured in the SMA, leading to a discrepancy with the experiment.

\subsection{Spin roton at integer filling fractions}

In order to gain simple physical understanding of the spin-roton minimum at $\nu=2/3$, we consider a simplified model system of a two component system at filling 2---i.e., an integer filling factor analog of the $\nu=2/3$ state. The  Hamiltonian is given by the following:
\begin{widetext}
\begin{align}
    &H=\sum_{n,m,\sigma}(n\hbar\omega+\sigma E_{z}/2) c_{nm\sigma}^{\dagger}c_{nm\sigma} +  \frac{1}{2}\sum_{\vec{q}}\sum_{n_{1,2,3,4}}\bar{V}_{n_{4},n_{3},n_{2},n_{1}}(\vec{q})  \left[\bar{\rho}_{n_{4}n_{1}}(-\vec{q})\bar{\rho}_{n_{3}n_{2}}({\vec{q}})-\delta_{n_{3},n_{1}}\bar{\rho}_{n_{4}n_{2}}(0)\right]\\
    &\bar{V}_{n_{4},n_{3},n_{2},n_{1}}(\vec{q})=V(\vec{q})F_{n_{4}n_{1}}(-\vec{q})F_{n_{3}n_{2}}(\vec{q})
\end{align}
\end{widetext}
where $c_{nm\uparrow/\downarrow}$ annihilates an spin-up/down electron in the $m$th guiding center and the $n$th Landau level. The form factor $F_{n'n}$ and the projected density operator $\bar{\rho}_{n'n}$ read
\begin{widetext}
\begin{align}
    F_{n'n}(\vec{q})&=\bra{n'}e^{-i\vec{q}\cdot{(-\vec{\Pi}}\times\hat{z})l_{B}^2}\ket{n}=\sqrt{\frac{2^{l'}l'!}{2^{l}l!}}(-iq_{x}+q_{y}\text{sgn}(n'-n))^{l-l'}L_{l'}^{l-l'}(\frac{\vec{q}^2}{2})e^{-\frac{q^2}{4}}\\
    \rho_{n'n}(\vec{q})&=\sum_{m',m,\sigma}\bra{m'}e^{-i\vec{q}\cdot{\vec{R}}}\ket{m}c_{n'm'\sigma}^{\dagger}c_{nm\sigma}
\end{align}
\end{widetext}
where $l\equiv\text{max}\{n,n'\},\ l'\equiv\text{min}\{n,n'\}$ and $L_{l}^{l'}(x)$ is the generalized Laguerre polynomial. We express the interaction Hamiltonian in terms of the projected density operator at the cost of an additional term proportional to $\bar{\rho}_{n'n}(0)$ to correct the ordering of fermion operators. This additional term cannot be dropped if the Hilbert space includes more than one Landau level since it in general does not commute with the first part of the interaction Hamiltonian. 
Let us calculate the dispersion of spin-flip collective modes within the Hilbert space spanned by $n=0,1$ LLs at $\nu=2$. When the exchange energy is the most dominant energy scale,
the ground state is a quantum Hall ferromagnet (QHF): a Slater determinant with electrons occupying the spin-up states of the $n=0$ and $n=1$ LLs.
Using the equation of motion describe in the last section, we arrive at the following eigenvalue equation:
\begin{widetext} 
\begin{equation} \label{eq:RPA}
    \sum_{n_{2},n_{1}=0}^{1}(\omega\chi_{n_{4}n_{3},n_{2}n_{1}}-\Omega_{n_{4}n_{3},n_{2}n_{1}}(\vec{q}) )Y_{n_{2}n_{1}}(\vec{q})=0
\end{equation}
where
\begin{align}
    \chi_{n_{4}n_{3},n_{2}n_{1}}&=\delta_{n_{4},n_{2}}\delta_{n_{3},n_{1}}\langle S_{n_{4}n_{3}}^{+}(\vec{q})S_{n_{2}n_{1}}^{-}(\vec{q})\rangle=\delta_{n_{4},n_{2}}\delta_{n_{3},n_{1}}N_{\phi}.\\
    \Omega_{n_{4}n_{3},n_{2}n_{1}}&=\langle[\bar{S}_{n_{4}n_{3}}^{+}(-\vec{q}),[H,\bar{S}_{n_{2}n_{1}}^{-}(\vec{q})]]\rangle\\
    &=\left[(n_{2}-n_{1})\hbar\omega+E_{z}\right]\chi_{n_{4}n_{3},n_{2}n_{1}}+\sum_{\vec{q},n'}\bar{V}_{n_{1}n'n_{4}n'}(\vec{q}')\delta_{n_{2}n_{3}}N_{\phi} -\sum_{\vec{q}}\bar{V}_{n_{1}n_{3}n_{2}n_{4}}(\vec{q}')\cos(\vec{q}'\wedge\vec{q})N_{\phi},\\
    \bar{S}_{n',n}^{-}(\vec{q})&=\sum_{m',m}\bra{m'}e^{-i\vec{q}\cdot{\vec{R}}}\ket{m}c_{n'm'\downarrow}^{\dagger}c_{nm\uparrow},
\end{align}
\end{widetext}
and $N_{\phi}$ is the number of flux quanta. To derive this equation, we have used the following properties of the quantum Hall ferromagnet:
\begin{align}
    &\bar{S}_{n'n}^{+}(\vec{q})\ket{\Psi_{QHF}}=0,\\ &\bar{\rho}_{n'n}(\vec{q})\ket{\Psi_{QHF}}=\delta_{\vec{q},0}\delta_{n'n}N_{\phi}.
\end{align}
At a given $q$, Eq.~\eqref{eq:RPA} is a $4\times 4$ matrix equation whose eigenvalues correspond to the four spin-flip collective modes plotted in Fig.~S8, where energy is measured relative to the Zeeman energy. Note there is always a mode with energy exactly equal to $E_{z}$ at $\vec{q}=0$ as required by the Larmor's theorem. Increasing the cyclotron energy will lower the energies of spin-flip transitions that lower the LL index.
This mode will mix with the Larmor mode at finite momentum $q=q^{*}$ and the consequent level-repulsion effect leads to a "spin-roton" minimum when $\hbar\omega_{c}$ is sufficiently large. 

\begin{figure*}
\begin{center}
\includegraphics[width=.8\textwidth]{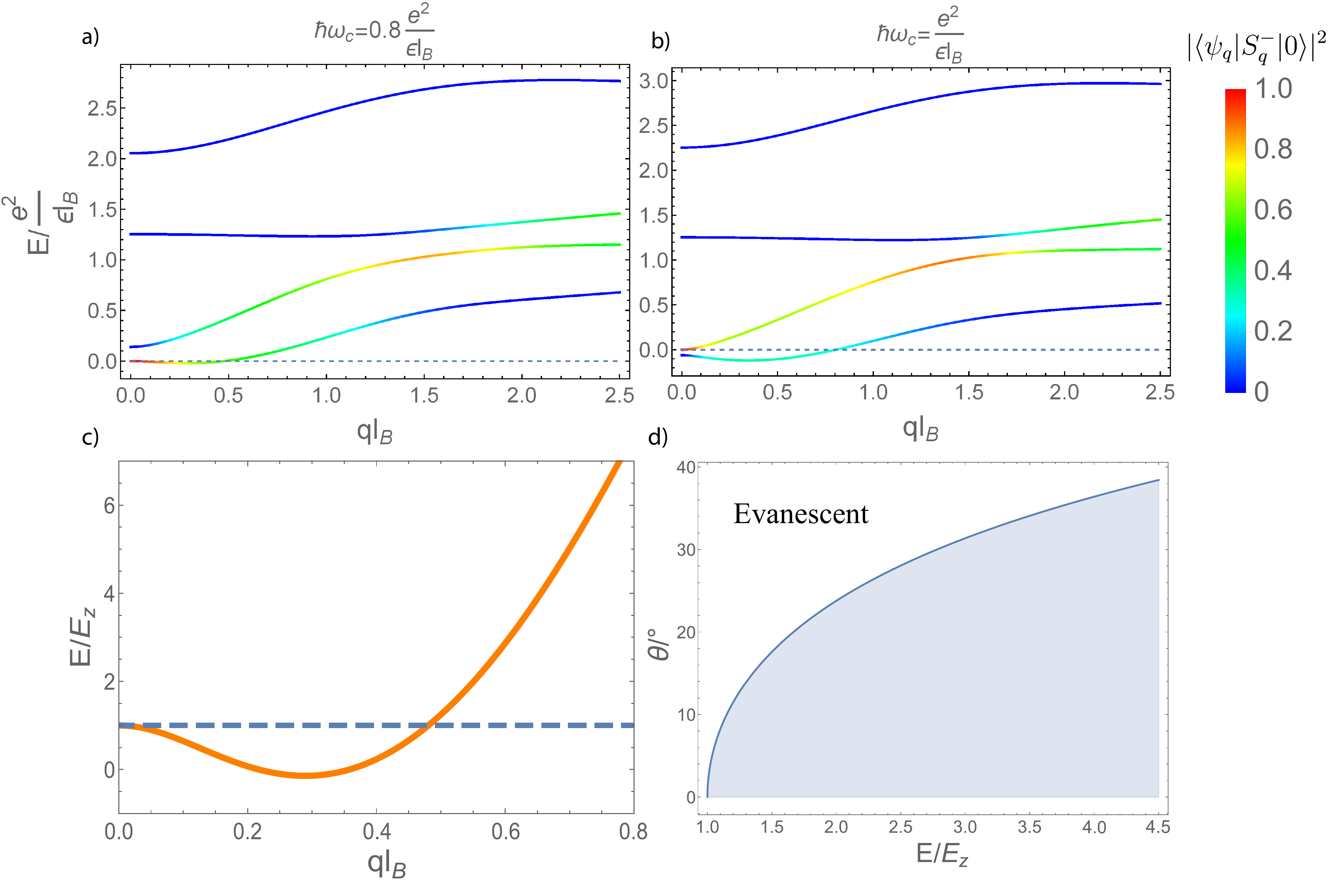}
\end{center}
\caption{The energy $E$ of the four spin-flip collective modes in the $N=0,1$ LL with the cyclotron energy a)$\hbar\omega_{c}=0.8e^2/\epsilon l_{B}$ and b)$e^2/\epsilon l_{B}$. The Zeeman energy is not included since it only produces a constant shift of all these dispersions. The line color stands for the overlap between the normal modes $\ket{\psi_{q}}$ and the single magnon wavefunction ansatz $S_{q}^{-}\ket{0}$, which is exactly one for the $q=0$ Larmor mode. c) The low energy dispersion of the lowest mode in a). Here we include the Zeeman energy. d)The phase space of magnon transmission in the incident angle - energy plane. When the incident angle from $\nu=2$ QHF onto the $\nu=1$  QHF becomes large, they become evanescent in the $\nu=1$ region.   
}\label{fig:spincurrent}
\end{figure*}

In this calculation, the ground state is always assume to be a QHF with maximum spin-polarization because the energy that favors a spin-polarized phase (exchange energy + Zeeman energy) is bigger than the energy that favors a spin-unpolarized phase (cyclotron energy), i.e.~$(\Delta_{ex}+E_{\rm Z})/\hbar\omega_c>1$. Note typically $\Delta_{ex}\gg E_{\rm Z}$.
As the ratio approaches one from above, $(\Delta_{ex}+E_{\rm Z})/\hbar\omega_c\rightarrow1^+$,
the spin-flip excitation spectrum of the QHF develops a deeper spin-roton minimum, as shown in Fig.~S8. This is because the energy of the spin-unpolarized state favored by $\hbar\omega_c$ becomes nearly degenerate to the ground-state energy of the QHF.  

At $\nu=2/3$, the Coulomb energy favors a spin-singlet state. When the Zeeman energy wins over the Coulomb energy, the ground state is spin-polarized and its spin-flip excitation also shows a shallow spin-roton minimum. The integer calculation suggests that this spin-roton minimum could arise from the competition between the spin-polarized state and the spin-singlet state. 
However, since the Zeeman energy merely provides a $q$-independent shift to the spin-flip spectrum, the characteristics of the spin-roton minimum (e.g.~position and depth) are determined solely by the Coulomb energy. 

Despite the difference between $\nu=2$ and $\nu=2/3$ excited wave-functions, their spin-roton dispersion is similar and this is sufficient to suppress magnon transmission into the stiffer $\nu=1$. By conservation of energy and transverse momentum $q_y$, we equate,
\begin{equation}
    \omega_{\text{rot}}(q_{x,in},q_y)=\omega_{\nu=1}(q_{x,out},q_y)
\end{equation}
where $\omega_{\nu=1}$ is the magnon dispersion of $\nu=1$ and $ \omega_{\text{rot}}$ is plotted in Fig.~S8c). Because $\omega_1$ is stiffer than $\omega_{\text{rot}}$, $q_{x,out}$ becomes imaginary when the incident angle $\tan^{-1}(q_y/q_{x,in})$ is large. The average transmission rate at low energy is thus reduced to less than $40\%$, already a large suppression compared to the 98\% transmission obtained by the naive SMA approximation. 

\subsection{Magnon Transmission at \texorpdfstring{$B>B_\perp^*$}{high field}}
For $B>B_\perp^*$, the high-field configuration $\ket{\Psi_H}$ is shown in Fig.~2d. We study magnon dispersion using the natural generalization of single-mode approximation. In metallic ferromagnets, magnons can be emitted when minority spin electron change to majority spin electron $e\downarrow\rightarrow e\uparrow$ and when majority spin hole change to minority spin hole $h\uparrow\rightarrow h\downarrow$. Similarly, magnons in $\ket{\Psi_H}$ can be created by $e\downarrow\rightarrow e\uparrow$ in the minority valley ($K'=B$) and $h\uparrow\rightarrow h\downarrow$ in the majority valley ($K=A$). Hence, the  spin-lowering process is in general a linear combination of the two valleys:
\begin{equation}
    O_{q}^{\dagger}=Y_{q}S_{qA}^{-}+Z_{q}S_{qB}^{-}.
\end{equation}

Using the equation of motion method, we arrived at the following:
\begin{equation}
    \hbar\omega_{q}\,\begin{pmatrix}\chi_{AA} & 0\\
0 & \chi_{BB}
\end{pmatrix}\cdot\begin{pmatrix}Y_{q}\\
Z_{q}
\end{pmatrix}=\begin{pmatrix}\Omega_{AA} & \Omega_{AB}\\
\Omega_{BA} & \Omega_{BB}
\end{pmatrix}\begin{pmatrix}Y_{q}\\
Z_{q}
\end{pmatrix}
\end{equation}
where
\begin{equation}
    \chi_{ij}=\langle S_{-qi}^{+}S_{qj}^{-}\rangle\;,\;\Omega_{ij}=\langle[S_{-qi}^{+},[H,S_{qj}^{-}]]\rangle,\qquad i,j=\{A,B\}
\end{equation}
and $\langle\rangle$ denotes the expectation value of the high field ground state $\ket{\Psi_{H}}$.
The static susceptibility can be easily evaluated to give
\begin{equation}
    \chi_{ii}=4(N_{i\uparrow}-N_{i\downarrow}).
\end{equation}
The frequency matrix is slightly more complicated. Using Eq.~\eqref{eq_commutators0z} and \eqref{eq_commutatorsxy}, we found:
\begin{widetext}
\begin{align}
\frac{\Omega_{AA}(q)}{\chi_{AA}} &=\Delta_{z}+\frac{2}{A}\sum_{k}(g_{o}(k)+g_{z}(k))\sin^{2}\left(\frac{k\wedge q}{2}\right)\left(-1+\mathcal{S}_{\nu}(k)\right)+\frac{u_{\perp}D}{2\chi_{AA}}\nonumber\\
&=\Delta_{z}-\rho_{\frac{1}{3}}q^2+O(q^4)\\
\frac{\Omega_{BB}(q)}{\chi_{BB}} & =\Delta_{z}+\frac{2}{A}\sum_{k}(g_{o}(k)+g_{z}(k))\sin^{2}\left(\frac{k\wedge q}{2}\right)+\frac{u_{\perp}D}{2\chi_{BB}}\nonumber\\
&=\Delta_{z}+u_{\perp}\frac{D}{2\chi_{BB}}+\rho_{1}q^2+O(q^4)\\
\frac{\Omega_{AB}(q)}{\chi_{AA}} &=-\frac{u_{\perp}D}{2\chi_{AA}}e^{-\frac{q^{2}l_{B}^{2}}{2}}\\
\frac{\Omega_{BA}(q)}{\chi_{BB}}&=-\frac{u_{\perp}D}{2\chi_{BB}}e^{-\frac{q^{2}l^{2}}{2}}
\end{align}
\end{widetext}
where we defined
\begin{align}
    D&=8(N_{A\uparrow}+(N_{A\uparrow}-N_{B\uparrow})-(N_{A\downarrow}-N_{B\downarrow})),\\
    \mathcal{S}_{\nu}(k)&=\frac{\langle\rho_{kA}\rho_{-kA}\rangle}{|N_{A\uparrow}-N_{A\downarrow}|}\equiv \frac{\langle(1+\tau^{z})_{-k}(1+\tau^{z})_{k}\rangle}{4|N_{A\uparrow}-N_{A\downarrow}|}
\end{align}
and also used the following properties of the $|\Psi_H\rangle$:
 \begin{equation}
    S_{q,A}^{-}|\Psi_H\rangle=S_{q,B}^{+}|\Psi_H\rangle=(\tau^{+}s^{-})_{q}|\Psi_H\rangle=(\tau^{-}s^{+})_{q}|\Psi_H\rangle=0,\quad \
\end{equation}
The normal mode frequency $\omega_{q}=\Delta_{z}+\rho_{2/3}^{(H)} q^2+O(q^4)$
\begin{align}
    \rho_{\frac{2}{3}}^{(H)}&=\frac{\rho_{1}-\rho_{\frac{1}{3}}}{2}+\frac{\rho_{1}+\rho_{\frac{1}{3}}}{2}\frac{\chi_{BB}-\chi_{AA}}{\chi_{BB}+\chi_{AA}}-\frac{u_{\perp}D}{2(\chi_{BB}+\chi_{AA})}\nonumber\\
    &\xlongequal[]{B=8T}1.9\rho_{1}
\end{align}
We see that the anitferromagnetic coupling between the spin in the two sublattices makes the magnon stiffer. The transmission probability, $T\approx 98\%$, of $\nu=1|2/3$ interface at high field is higher than the low field.

\end{document}